\documentclass[preprint,amsmath,amssymb,aps,pre]{revtex4-2}
\usepackage{graphicx}
\newcommand{\df}{\mathrm{d}}
\begin{document}
	\title{Dynamic magnetic response in ABA type trilayered systems and compensation phenomenon}
	\author{Enakshi Guru}
	\email{enakshiguru@sncwgs.ac.in}
	\author{Sonali Saha}
	\email{sonali@sncwgs.ac.in}
	\affiliation{Sarojini Naidu College for Women\\ 30, Jessore Road, Dumdum, Kolkata, India.}
	\author{Sankhasubhra Nag}
	\email{sankha@rkmvccrahara.ac.in}
	\affiliation{Ramakrishna Mission Vivekananda Centenary College\\ Rahara, Kolkata, India.}
	\begin{abstract}
		Dynamic magnetic response in a trilayered structure with non-equivalent layers (${\bf ABA}$ type) has been studied with Monte Carlo simulation using Metropolis algorithm. In each layer, ferromagnetic (FM) nearest neighbour Ising interactions are present along with antiferromagnetic (AFM) nearest neighbour coupling across different layers. The system is studied under a harmonically oscillating external magnetic field. It is revealed that along with dynamic phase transition (DPT), compensation phenomenon emerges in this system under dynamic scenario too. This feature in dynamic case is unique for such trilayered systems only, in contrast to the bulk system reported earlier. The temporal behaviour of the magnetisation of each individual layer shows that different magnetic response of the non-equivalent layers  results into such dynamic compensation phenomenon. The difference in response also results into warping of the dynamic hysteresis loops, under various external parameter values, such as amplitude of the oscillating field and temperature.
	\end{abstract}
	\maketitle
	\section{Introduction}
	In the area of non-equilibrium phenomenon, dynamic phase transition (DPT) attracted attention of a lot of physicists (e.g.refs ~\cite{Chakrabarti1999,Acharyya2005,Yueksel2022} and the references therein.).  In  1990, DPT in magnetic systems was studied by Tome and Oliviera using mean field approach with classical ferromagnetic Ising interactions among spin sites in a sinusoidally time varying but spatially uniform magnetic field and they found a phase transition using average value of magnetisation ($M$) as order parameter~\cite{Tome1990}.
	
	Later in 1993, the frequency response of the hysteric nature of such time varying (2--4 dimensional) systems under external oscillatory field was studied using Monte-Carlo (MC) simulations and compared the results with those found from mean field theory~\cite{Acharyya1993}. With the variation of frequency ($ f $) and amplitude of the external field ($ B_0 $) at different temperatures ($ T $), corresponding effect on area of the dynamic hysteresis curve was observed. Apart from it, in the same work using the time averaged magnetisation of the system as order parameter, dynamic phase transition was investigated and phase boundary in $ B_0-T $ plane was identified. In 1995, using both MC simulations and mean field calculations, these studies were extended with an additional transverse field to find the scaling form and the corresponding parameters (in 1--4 dimensions)~\cite{Acharyya1995}. They also explored the relations between susceptibilities and effective time delay in the magnetic response of the system. Later Chatterjee \& Chakrabarti investigated the effect of a single square pulse of external magnetic field on an ordered Ising spin system~\cite{Chatterjee2004}.
	
	In 2001, DPT was studied under oscillatory external magnetic field using MC simulations in thin ferromagnetic (FM) film modelled by classical Heisenberg interactions~\cite{Jang2001}; with bilinear anisotropy and competing static surface fields. The phase boundaries in this case was depending upon $ B_0 $  and $ f $ of the oscillation. Later same sort of studies were repeated with pulsed oscillatory external field~\cite{Jang2003}. 
	
	The dynamic responses of Ising metamagnet (cubic structure with FM interactions within a layer and AFM interlayer interactions) in the presence of a
	sinusoidally oscillating magnetic field were studied by MC simulation~\cite{Acharyya2011}. The time averaged staggered magnetisation assumed the role of dynamic order parameter there.  Later the dynamic response of a two dimensional Ising ferromagnet to a plane polarised standing magnetic field wave was modelled and studied by MC simulation in two dimensions~\cite{Halder2016}. Two main dynamical phases namely, pinned and oscillating spin clusters, were detected. The dynamic response of spin-$ S $ (with $ S=1, 3/2, 2, 3 $) Ising ferromagnet to the plane propagating wave, standing magnetic field wave and uniformly oscillating field with constant frequency were studied~\cite{Halder2017} separately in two dimensions by extensive MC simulations. Depending upon the strength of the magnetic field and the value of spin state in the Ising spin lattice two different dynamical phases were observed. Very recently, studies on two dimensional square lattice of spins with X-Y interactions under periodic square waves, revealed four dynamic phases by mean field approach as well as by MC simulations~\cite{Pal2024}.

	On the other hand, multilayered systems under equilibrium conditions, also drew  attention from the scientific community. A comprehensive experimental study on the AFM interlayer exchange coupling in high quality epitaxial all-semiconducting EuS/ PbS/ EuS trilayers was reported back in 2004 ~\cite{Smits2004}.  In 2012, Szalowski \& Balcerzak studied multilayered spin-1/2 anisotropic classical Heisenberg systems with distinct intra- and inter-layer interactions under constant external magnetic field~\cite{Szalowski2012}. In the next year, Yadari and others performed investigations on Ising super-lattices using MC simulations under static magnetic field to observe thermal behaviour of magnetisation, that of susceptibilities, etc.~\cite{Yadari2013}.  Naji and other workers performed studies on trilayer super-lattice with spin-1/2 , 1 and 3/2 arranged in layers, with open and periodic boundary conditions using mean field approach~\cite{Naji2014a} and MC simulations~\cite{Naji2014}. They explored corresponding phase diagram and variation of $M$ and susceptibilities with temperature considering nearest neighbour interaction and static external field.

	In 2017, phase diagrams using mean field theory and effective field theory for trilayered spin-1/2 systems were studied~\cite{Diaz2017} and found the dependence of compensation temperature and critical temperature on the system parameters in the Hamiltonian. 
	Cubic multilayer systems which consists of non-equivalent layers (with and without dilution in one type of layer) were studied~\cite{Diaz2018} using MC simulations and compared the results with those obtained from different approximations. Further in 2019 Diaz \& Branco~\cite{Diaz2019}, using MC methods, investigated  trilayered systems (of {\bf AAB} and {\bf ABA} type) with FM intra-layer and AFM interlayer coupling. The compensation phenomenon was verified in this case; the compensation and critical temperatures as functions of the Hamiltonian parameters were obtained. Sajid and Acharyya subsequently studied effects of random non-magnetic impurity in the trilayer systems~\cite{Sajid2020} and obtained phase diagram for selected Hamiltonian parameters and impurity concentrations in {\bf ABA} type. Later the trilayered Ising systems (of both {\bf ABA} and {\bf AAB} type) were investigated with variation of interlayer coupling ~\cite{Chandra2021}, and with site dilution in all layers ~\cite{Chandra2023a}. The investigation for compensation effect was  extended for additional next nearest neighbour intra-layer interaction in {\bf ABA} type of such trilayer systems ~\cite{Guru2022}. In the above mentioned works on trilayered {\bf ABA}, magnetic responses at equilibrium conditions were studied. Few works were done on the effect of external field randomly varying in time (uniformly distributed~\cite{Chandra2021a} or in Gaussian distribution~\cite{Chandra2023}) on equilibrium state at different temperature. The fluctuations in  magnetisation due to such disturbance on the previously achieved equilibrium state  were studied.

	In the above mentioned works on trilayered systems, with FM intra-layer and AFM inter-layer interactions (both of Ising type), distinctive behaviours such as compensation phenomenon with variety of features were found. It is due to the interplay between FM and AFM interactions and due to different magnetic responses from different layers. Thus it is expected that this type of systems may show certain analogous characteristic features in DPT too; which is absent in the case of bulk. In our present work, DPT in {\bf ABA} type of trilayer system is explored; where all the layers consist of spins interacting by intra-layer FM ising coupling and AFM interactions across adjacent different layers.  The middle layer (type {\bf B}) is different in the sense that the Ising interaction strength within this layer is double of that in adjacent layers (type {\bf A}).
	
	\section{Model and Methodology}
\begin{figure}[h]
		\centering
		\includegraphics[width=.5\linewidth]{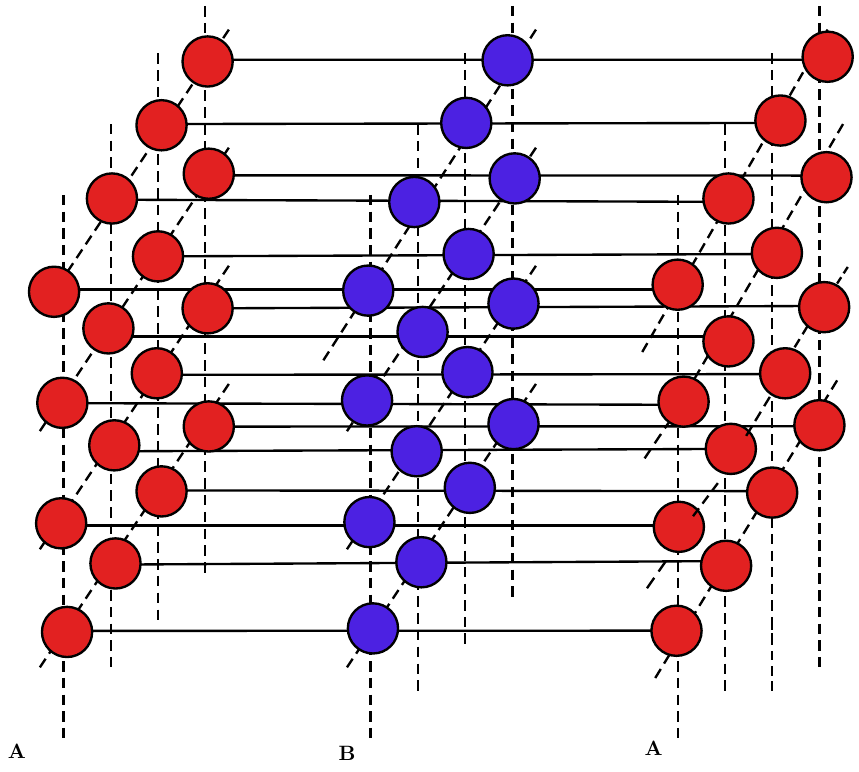}
		\caption{
			Schematic representation of a portion of  the square mesh like three layers of the {\bf ABA} system is displayed (see text). The side layers are composed of $A$ type sites, marked with red colour (online version) and the middle layer is composed of  $B$ type sites, marked with blue colour (online version). The NN intra-planer  FM interactions ($J_{nn}$) are shown with short dashes (-)  while solid lines are used for the inter-planer AFM interactions ($J_{n,n+1}$).}
		\label{fig:fig1}
	\end{figure}

	In this current work the investigation is done on a trilayered system in an external time varying (sinusoidally) but spatially uniform magnetic field. The trilayer is of {\bf ABA} type i.e., two outer layers (see figure~\ref{fig:fig1}) are of similar type while the middle one is different. Hence the Hamiltonian of the trilayered system with nearest neighbour (NN) Ising interactions in presence of external oscillating magnetic field  is:
	\begin{align}\label{key}
		H_E= -\sum_{n=1}^{3} J_{nn} \sum_{i,j=1}^{L}s_{i,j}^{(n)}. \left(s_{i+1,j}^{(n)}+ s_{i,j+1}^{(n)}\right)
		-\sum_{n=1}^{2}J_{n,n+1}\sum_{i,j=1}^{L}&\left(s_{i,j}^{(n)}.s_{i,j}^{(n+1)}\right)\nonumber\\ -&B_0\cos(2\pi ft)\sum_{n=1}^{3}\sum_{i,j}^{L} s_{i,j}^{(n)},
	\end{align} ($ s_{i,j}^{(n)} $ being the spin value at the lattice site located at $ j $'th column and $ i $'th row in the $ n $'th layer). Here the first term with summation stands for all spin-spin intra-planer interactions  within all $n$'th layers ($ n=1,2,3 $). The second summation term indicates the contributions from inter-planer NN interactions and the third one represents onsite spin-field interaction. Obviously $ J_{nn} $ stands for intralayer interaction strength and $ J_{n,n+1} $ represents the interlayer one. Here $J_{11}=J_{33}=J_{AA}$ and $J_{22}=J_{BB}$, while $J_{12}=J_{23}=J_{AB}$; where $J_{AA}$, $J_{BB}$ and $J_{AB}$ are constants. 
	
	In these numerical studies on the trilayered system, each layer is a two dimensional square lattice and is composed of $100\times 100$ spins (i.e. $ L=100 $). Here classical Ising model is used, where $s_{i,j}^{(n)}=\pm 1$. Within the plane, the spin-spin interactions satisfy the periodic boundary conditions, whereas the system is open out of the plane.  The positive values of the interaction coefficients $J_{nn}$'s indicate FM interactions strength  whereas negative values indicate AFM interactions; while the same is true for $ J_{n,n+1} $, the coefficient for interlayer interaction strength. To study the effect of external oscillating field  on this trilayered system we have investigated with AFM interlayer interactions.
	Unlike the previous studies on trilayer (see refs.~\cite{Naji2014a, Naji2014,Diaz2017, Diaz2019, Chandra2020, Chandra2021,Chandra2023a,Guru2022} etc.), here the system is far from equilibrium due to the presence of oscillatory external field. The scope of Metropolis algorithm has been applied ~\cite{Stauffer1989, KurtBinder2002} to study the dynamical behaviour of the system. 
	
	In the MC simulation for this work, the initial configuration has been chosen a random state. In every step a single site is selected randomly and flipped. The new spin configuration was accepted with the probability $p=\min\left[1,\exp(-\Delta E/k_BT)\right]$;
	where $\Delta E$ is the change in energy due to spin flip.  Such $3\times L\times L$ number of spin flips is referred as one Monte Carlo step per spin (MCSS), which is the unit of time (in our case it is $10^{-5}$). This choice is arbitrary and not related to real time. Here it has been selected such that at our highest frequency of applied field (i.e for $f=100$) the system can go through 1000 MCSS within one complete cycle. The result was analysed taking the average over five such complete cycles. To study the effect of temperature, the system was allowed to cool slowly, at each step the cooling rate was $0.005$ in the unit of $ J_{BB}/k_B $. 
	
	\section{Results and Discussion}
	The primary objective of this study is to investigate how  {\bf ABA} trilayered systems behave in  presence of a sinusoidally oscillating external field and also to compare the observed behaviour with those of a bulk system. In connection with that, it is relevant to investigate the nature of dynamic phase transition that may happen in this case. As already used by various authors in the case of bulk systems, here too, the time averaged magnetisation $ Q $ ($ =\frac{1}{\tau}\int_0^{\tau}M\df t  $; where $ \tau=1/f $) is chosen as the order parameter. Accordingly, the time variation of $ M $ for different amplitudes ($ B_0 $, with same unit as that of $ J $) of the oscillating field is studied  and that too along with the variation of temperature. Magnetisation ($ M $) is the total spin of the system per site, i.e., \[
		M=\frac{\sum_{n=1}^{3}\sum_{i,j=1}^{L}s_{i,j}^{(n)}}{3L^2}.
	\]
	
\begin{figure}[h!]
		\centering
		\includegraphics[width=0.5\linewidth]{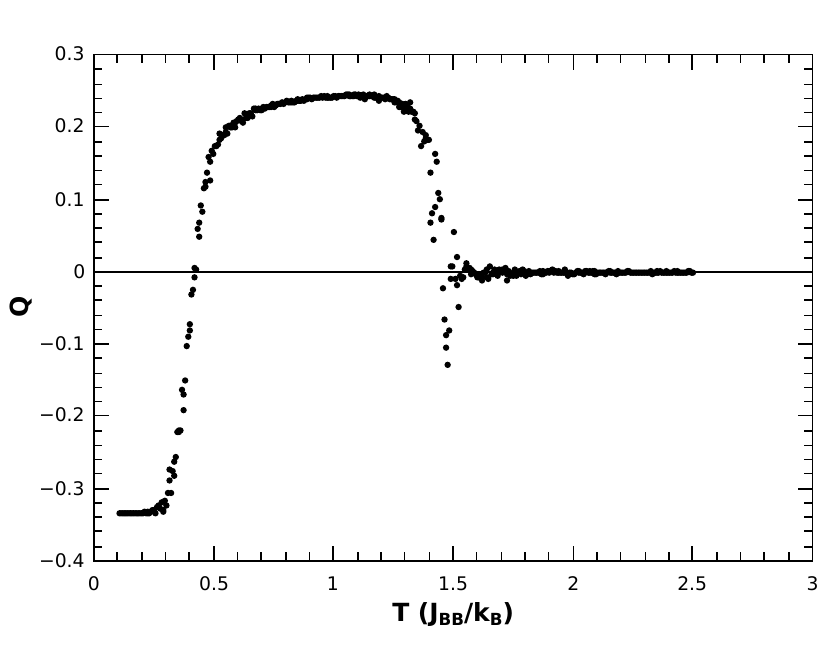}
		\caption[fig2]{Temperature variation of dynamic order parameter $Q$, for $B_0=0.5$ and $f=100$.}
		\label{fig:fig2}
	\end{figure}

\begin{figure}[h]
		\centering
		\includegraphics[width=0.7\linewidth]{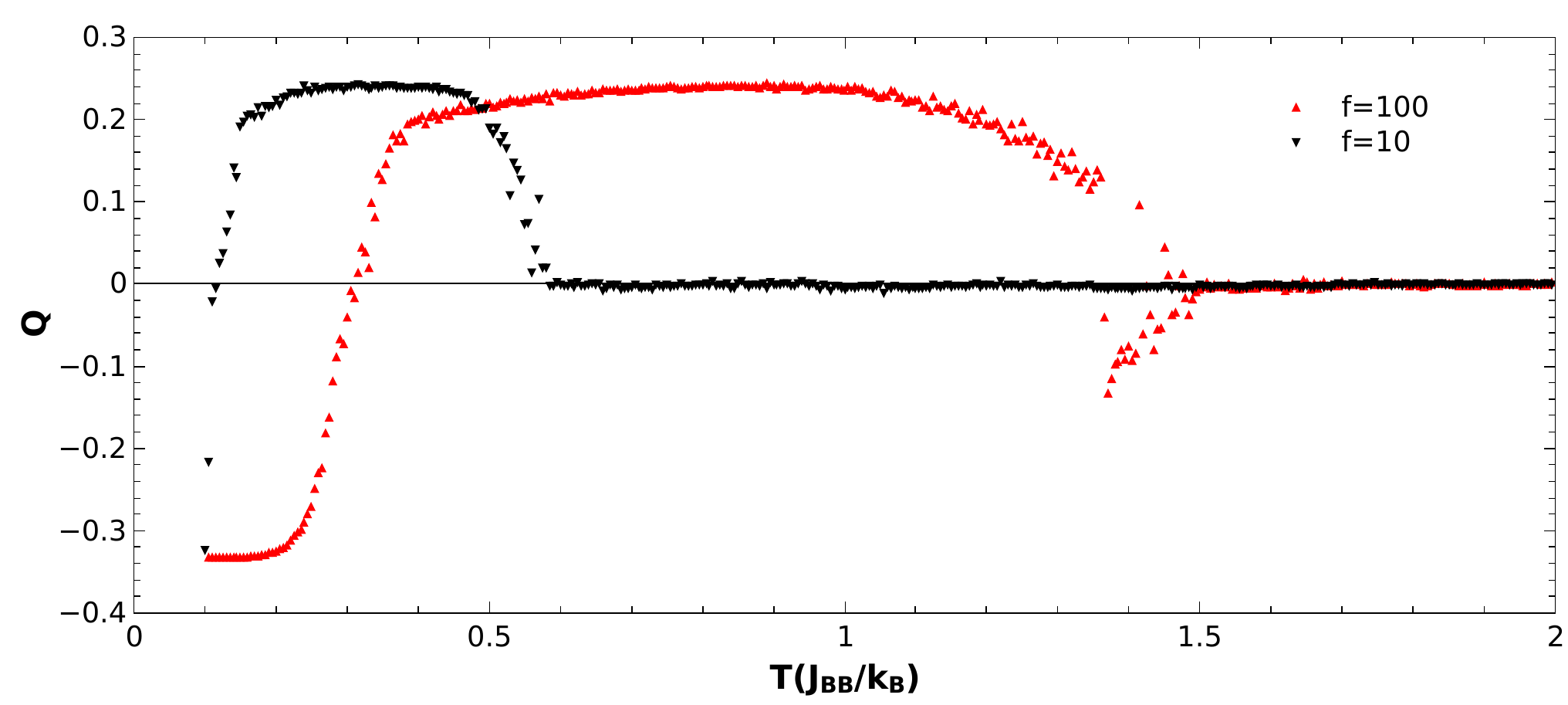}
		\caption[fig3]{Comparison of temperature dependence of dynamic order parameter $Q$ at two different frequencies (see legends). $B_0=0.5$ for both the cases.}
		\label{fig:fig3}
	\end{figure}

	In contrast to the bulk system, the most startling feature observed for this trilayered structure is the following behaviour. Here at high temperature region, higher than a critical one which may be termed as dynamic transition temperature ($ T_{dt} $), $ Q $ vanishes (figure~\ref{fig:fig2}).
	It is a common feature for DPT in bulk systems too (see e.g. refs.~\cite{Tome1990,Acharyya1993,Acharyya1995,Chakrabarti1999}). But in this case for lower values of $ B_0 $, when the temperature is less than $ T_{dt} $, $M$ becomes non-zero with non-monotonic behaviour as a function of temperature and vanishes again at an intermediate value. That temperature may be called as dynamic compensation temperature ($ T_{dcom} $), in  analogy with the term `compensation temperature'  used in literature (e.g. refs.~\cite{Diaz2017,Diaz2019,Sajid2020,Chandra2021} ) for equilibrium phase transition and compensation phenomenon for trilayered systems. If the temperature is further lowered down $ Q $ again rises up with opposite sign. The feature remains qualitatively same for different frequencies of the external field $ B $ (figure~\ref{fig:fig3}); though for lower frequencies both $ T_{dt} $ and $ T_{dcom} $ shift to the lower end of temperature and their differences also diminish.  On the other hand, with increase of $ B_0 $ the transition temperature $ T_{dt} $, moves towards lower end along with $ T_{dcom} $ (not shown here).
	
	To investigate the mechanism behind this feature of time averaged magnetisation $ Q $, the temporal behaviour of $M$ for this trilayer and those for individual ones separately, are studied. For the sake of discussion one may divide the entire range of temperature into following three zones: (I) $ T>T_{dt} $, (II) $ T_{dcom} < T <T_{dt} $ and (III) $ T<T_{dcom} $; for each zone (for frequency value 100) one representative value is chosen: $ T=1.8 $ for zone (I), $ T=0.8 $ for (II) and $ T =0.2 $ for (III).
	\begin{figure}[h]
		\centering
		\includegraphics[width=0.8\linewidth]{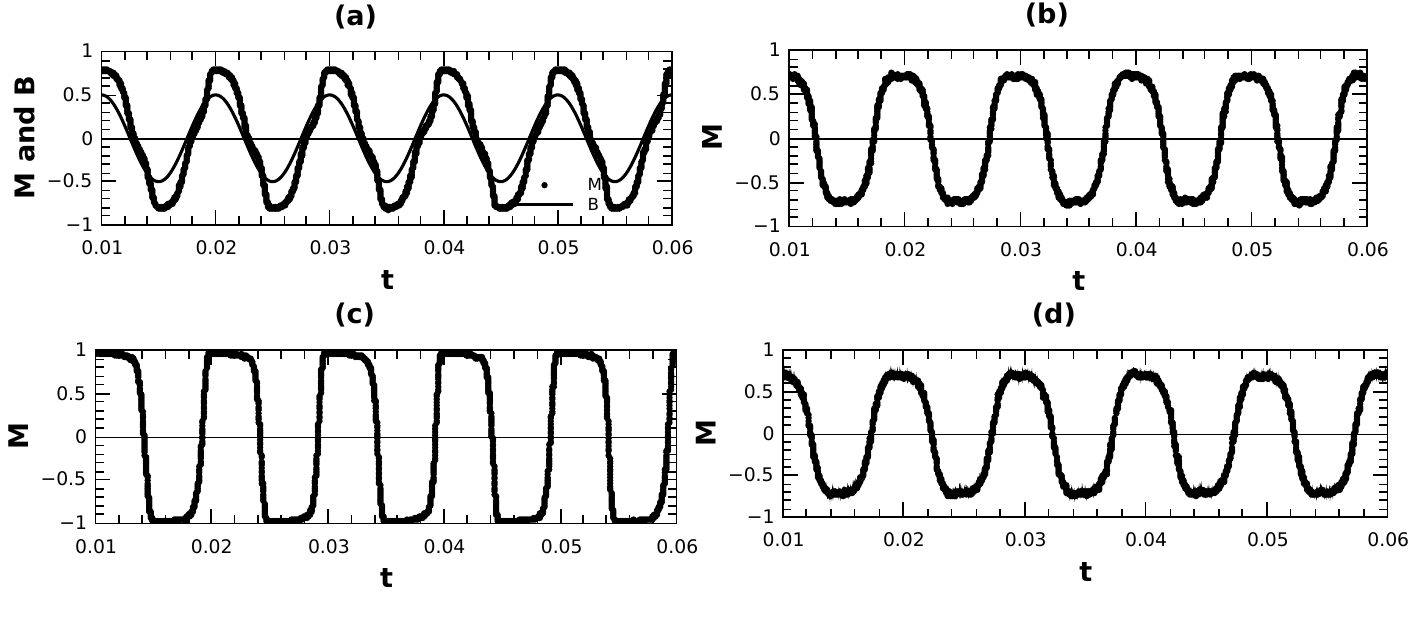}
		\caption[fig4]{(a) Time variation of $ M $ of the system and that for the external field $ B $ (see legends) at $T=1.8$ with $B_0=0.5$ and $f=100$. The same for individual side layers, in (b) and (d), for the middle layer in  (c), are shown.}
		\label{fig:fig4}
	\end{figure}
	
	The value of $ B_0=0.5 $ is initially chosen to focus on the region where non-monotonic behaviour of $ Q $ (with $ T $) is predominant. At a temperature (represented by $ T=1.8 $) in zone (I),   total magnetisation  oscillates symmetrically about zero, within a range $\approx\pm 0.8  $ with a distorted sinusoid wave form (see figure~\ref{fig:fig4} (a)). Hence $ Q $ becomes zero. The distortion comes due to the different magnetic response of the middle layer and those of side layers (figure~\ref{fig:fig4}(b),(c),(d)). 		
	
\begin{figure}[h]
		\centering
		\includegraphics[width=0.8\linewidth]{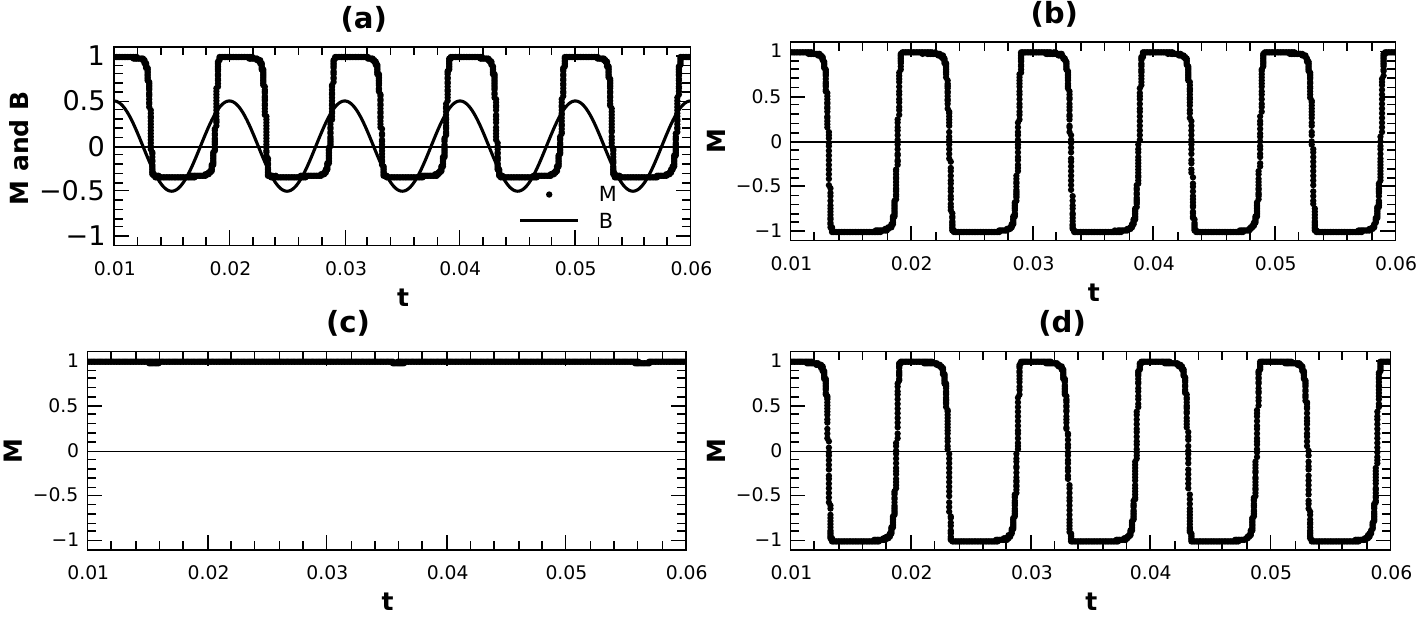}
		\caption[fig5]{(a) Time variation of $ M $ of the system and that for the external field $ B $ (see legends) at $T=0.8$ with $B_0=0.5$ and $f=100$. The same for individual side layers, in (b) and (d), for the middle layer in  (c), are shown.}
		\label{fig:fig5}
	\end{figure}

	For the intermediate zone (II) (represented by $ T=0.8 $), total magnetisation toggles  asymmetrically about zero in between $ +1 $ and $ -0.33 $ (figure~\ref{fig:fig5}). Hence as a result, $ Q $ becomes positive. Studies on individual layer reveals that the middle layer gets pinned along one direction (with $M=+1 $) throughout a cycle; whereas that for side layers toggles symmetrically (figure~\ref{fig:fig5}(b), (c), (d)), giving rise to the asymmetry in $ M $ for the whole system. 
	
\begin{figure}[h]
		\centering
		\includegraphics[width=0.6\linewidth]{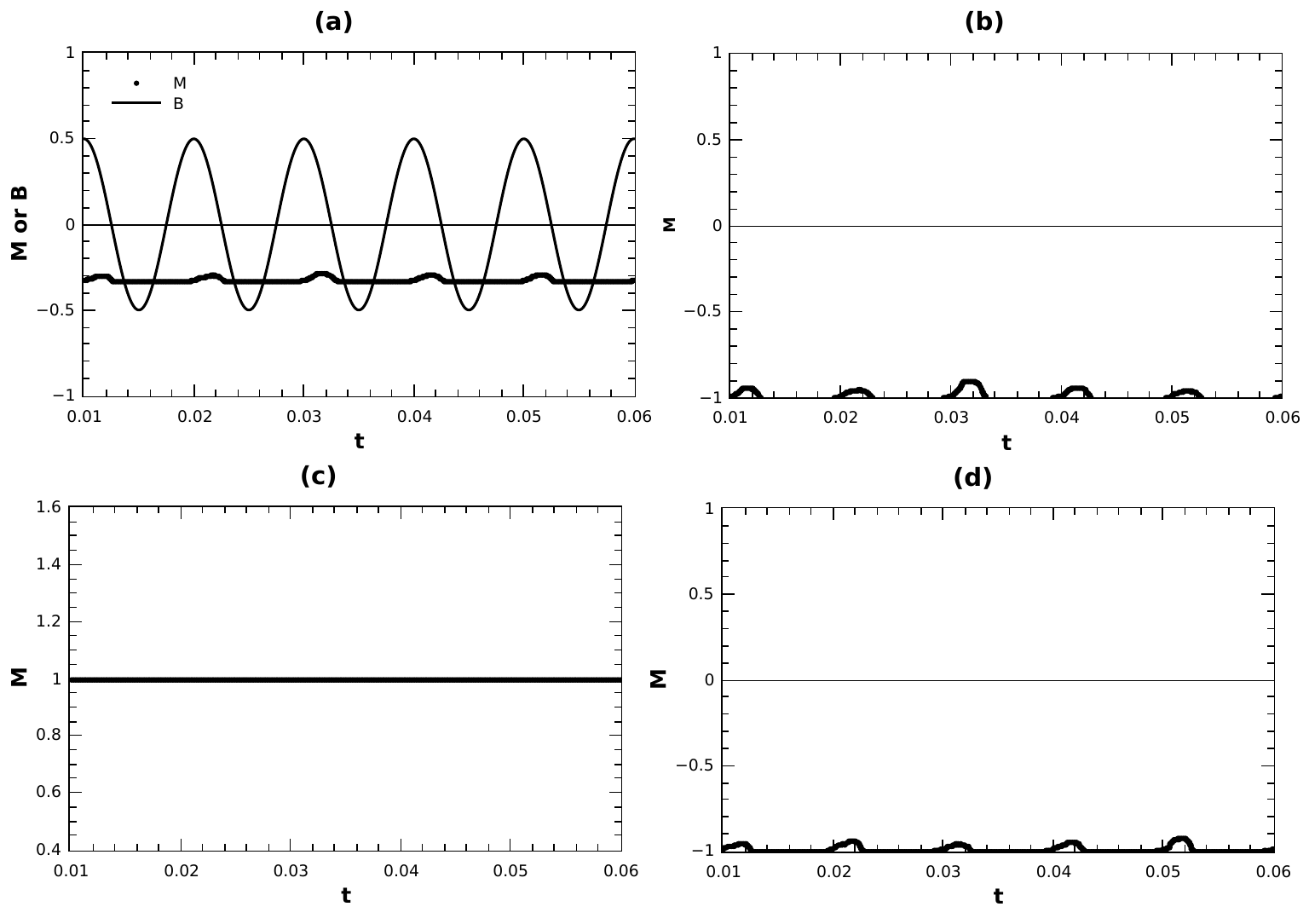}
		\caption[fig 6]{(a) Time variation of $ M $ of the system and that for the external field $ B $ (see legends) at $T=0.2$ with $B_0=0.5$ and $f=100$. The same for individual side layers, in (b) and (d), for the middle layer in  (c), are shown.}
		\label{fig:fig6}
	\end{figure}

	Once temperature decreases below $ T_{dcom} $ in zone (III) (represented by $ T=0.2 $), the side-layers in most of the time within a cycle settles in opposite direction with $ M=-1 $ (figure~\ref{fig:fig6}), turning the value of $Q$ in the opposite direction indicated by its negative value. When the temperature is lowered down from $ T_{dt} $, for the intermediate values of temperature, $M$ of the side-layers begins to oscillate asymmetrically with a bias towards $ M=-1 $. Hence the upper-bound of oscillating total magnetisation, diminishes progressively from $ +1 $ (not shown here), which in turn diminishes the value of $ Q $ (see figure~\ref{fig:fig3}). At $ T_{dcom} $, the break-even situation arises when the average magnetisation of middle layer  gets exactly cancelled or compensated by the same of the side-layers in opposite direction; hence the coinage of the word `compensation' comes. 

\begin{figure}[h!]
		\centering
		\includegraphics[width=0.7\linewidth]{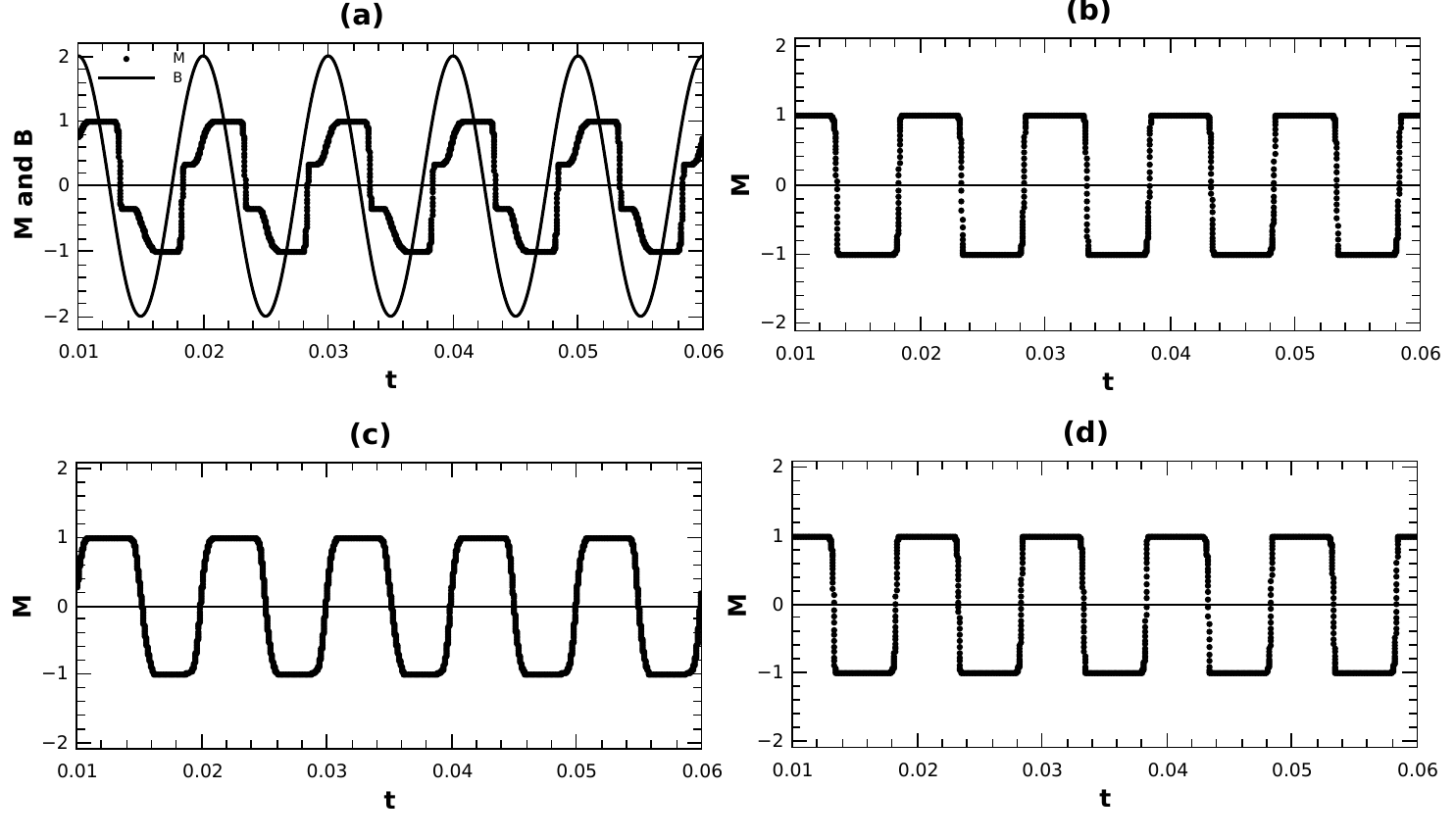}
		\caption[fig 7]{(a) Time variation of $ M $ of the system and that for the external field $ B $ (see legends) at $T=1.8$ with $B_0=2$ and $f=100$. The same for individual side layers, in (b) and (d), for the middle layer in  (c), are shown.}
		\label{fig:fig7}
	\end{figure}

	For sufficiently higher values of the amplitude of external field the asymmetry in oscillations of magnetisation for all the layers withers away even for the lower values of temperature (figure~\ref{fig:fig7}); thus $ Q $ vanishes. But for those cases too, the difference in magnetic responses becomes apparent in the phases of their oscillations (of magnetisation). This difference results into intermediate steps in toggling of $M$ between $ \pm 1 $.  
	
	In this system, the interplay between the layers {\bf A} and {\bf B}, plays a crucial role in determining the behaviour. The inner layer of type {\bf B} is much more resilient to the change in external field due to its larger intra-layer  coupling strength compared to type {\bf A} layer; it tries to retain their internal ordering to a greater extent. Moreover due to the AFM interlayer interactions, spins in side-layers of type {\bf A} are prone to align themselves in the opposite direction to those of the middle layer (type {\bf B}). Hence in presence of a specific value of  external field, spins from all the layers tend to align themselves along the field (even destroying their internal ordering); but those in the side layers experience a competition between the tendency to align themselves along the external field and that to anti-align with respect to their counterpart in the inner (middle) layer. Due to both the factors the internal ordering within the side-layers is more fragile compared to that of middle layer. This explains the comparatively ready response of the side layers to the change in external field in time-varying  scenario; which in turn justifies the difference in dynamic responses between middle and side layers of this trilayered structure. In the results discussed in previous paragraphs, it is this difference that plays the crucial role behind the compensation phenomenon in this dynamic case.
	
\begin{figure}[h!]
		\centering
		\includegraphics[width=0.5\linewidth]{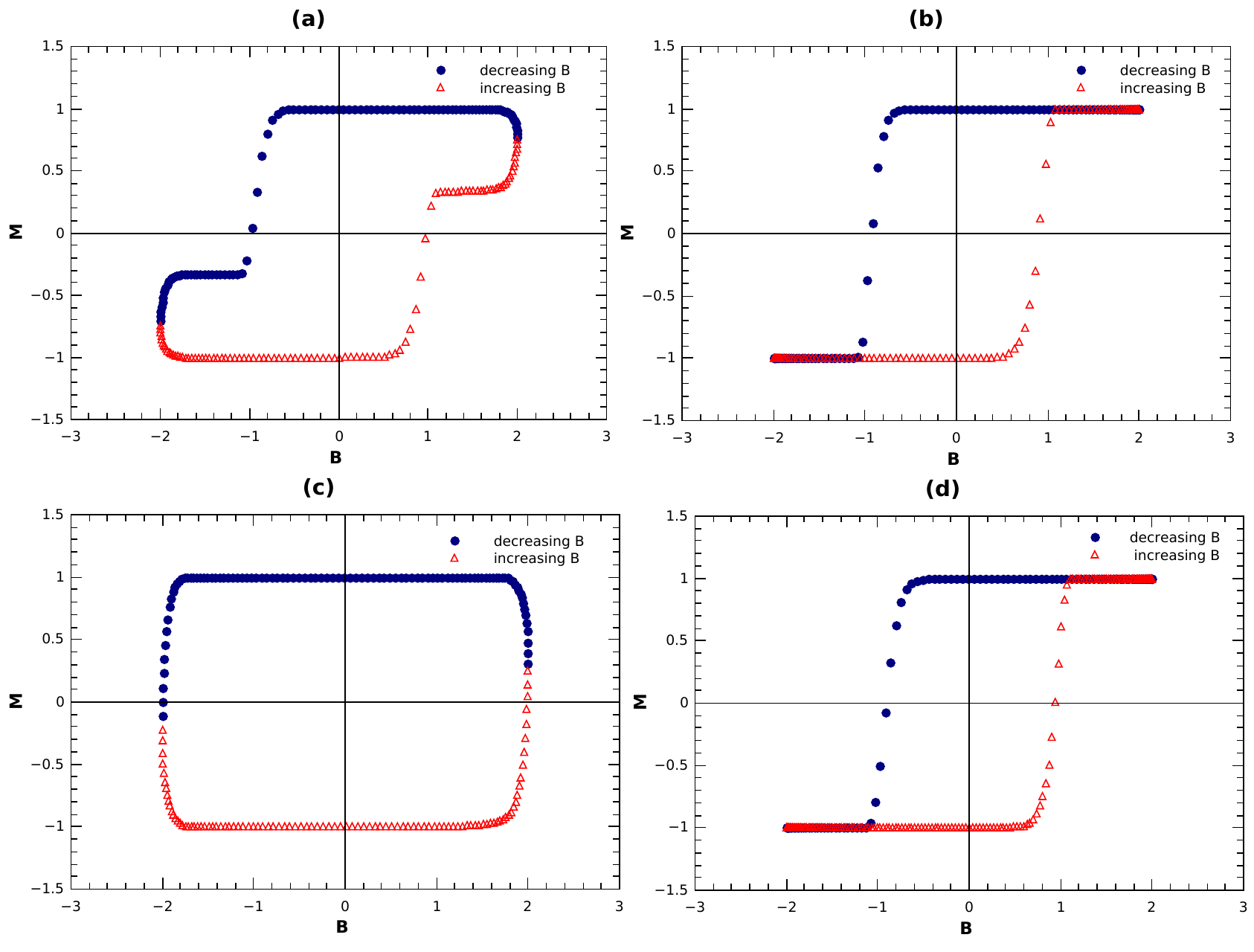}
		\caption[fig 8]{(a) Dynamic hysteresis loop of the trilayered system at $T=0.2$  for $B_0=2$ and $f=100$. Half of a cycle with decreasing $B$ and that for increasing $B$ are shown with separate symbols (see legend).  The same for the individual side layers are shown in (b) \& (d) and for middle one in (c).}
		\label{fig:fig8}
	\end{figure}

	This variation of magnetic response becomes even more apparent from the dynamic hysteresis curve, where for larger values of $ B_0 $ (e.g. at $ B_0=2 $), the delayed response of middle layer produces larger loop compared to those produced by side ones, making the loop for the whole structure (essentially the superposition of those for all three layers) symmetric but warped (figure~\ref{fig:fig8}). In certain cases with high values of $ T $ and $ B_0 $, it is so warped that instead of a single loop three loops are formed within a cycle (not shown here). 
	\begin{figure}[h!]
		\centering
		\includegraphics[width=0.7\linewidth]{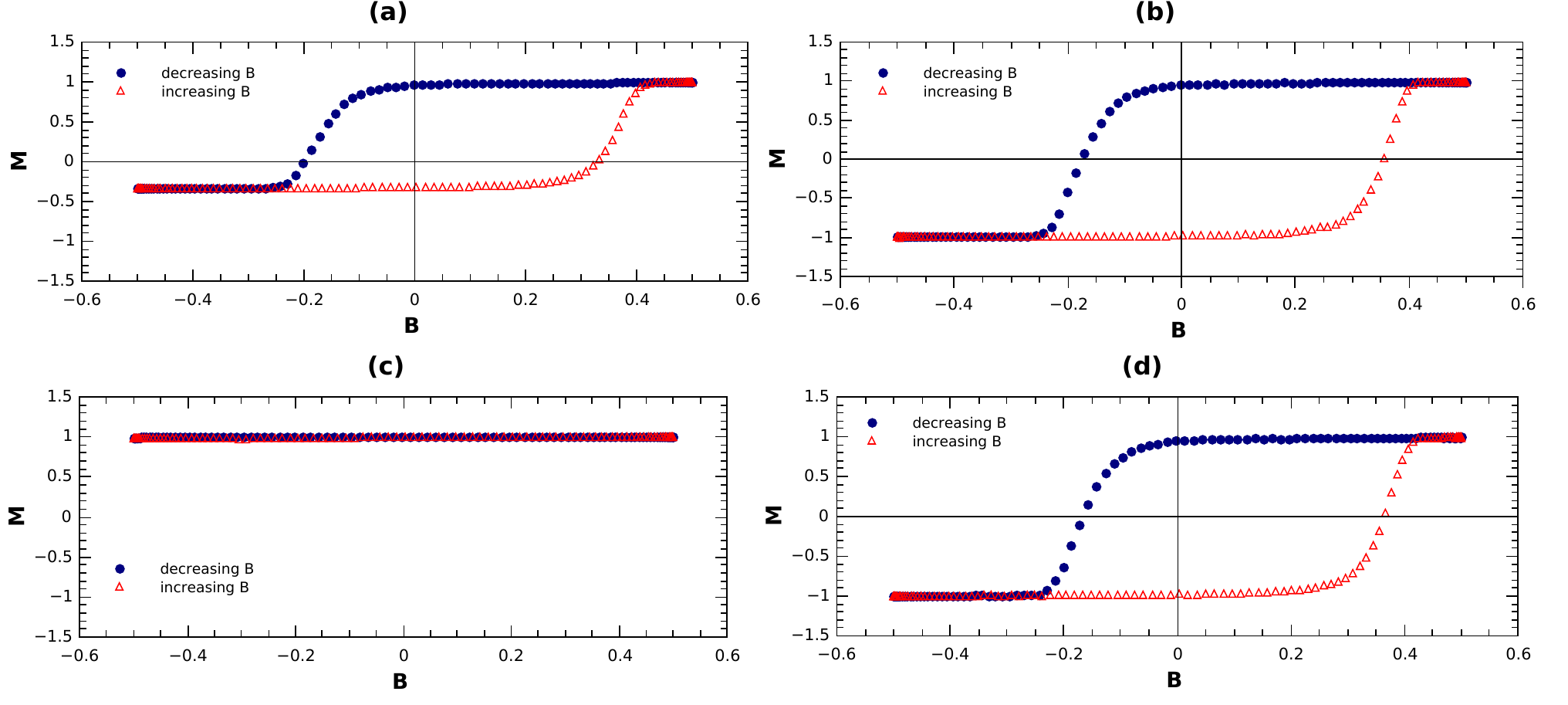}
		\caption[fig 9]{(a) Dynamic hysteresis loop of the trilayered system at $T=0.8$  for $B_0=0.5$ and $f=100$. Half of a cycle with decreasing $B$ and that for increasing $B$ are shown with separate symbols (see legend).  The same for the individual side layers are shown in (b) \& (d) and for middle one in (c).}
		\label{fig:fig9}
	\end{figure}
	\begin{figure}[h!]
		\centering
		\includegraphics[width=0.7\linewidth]{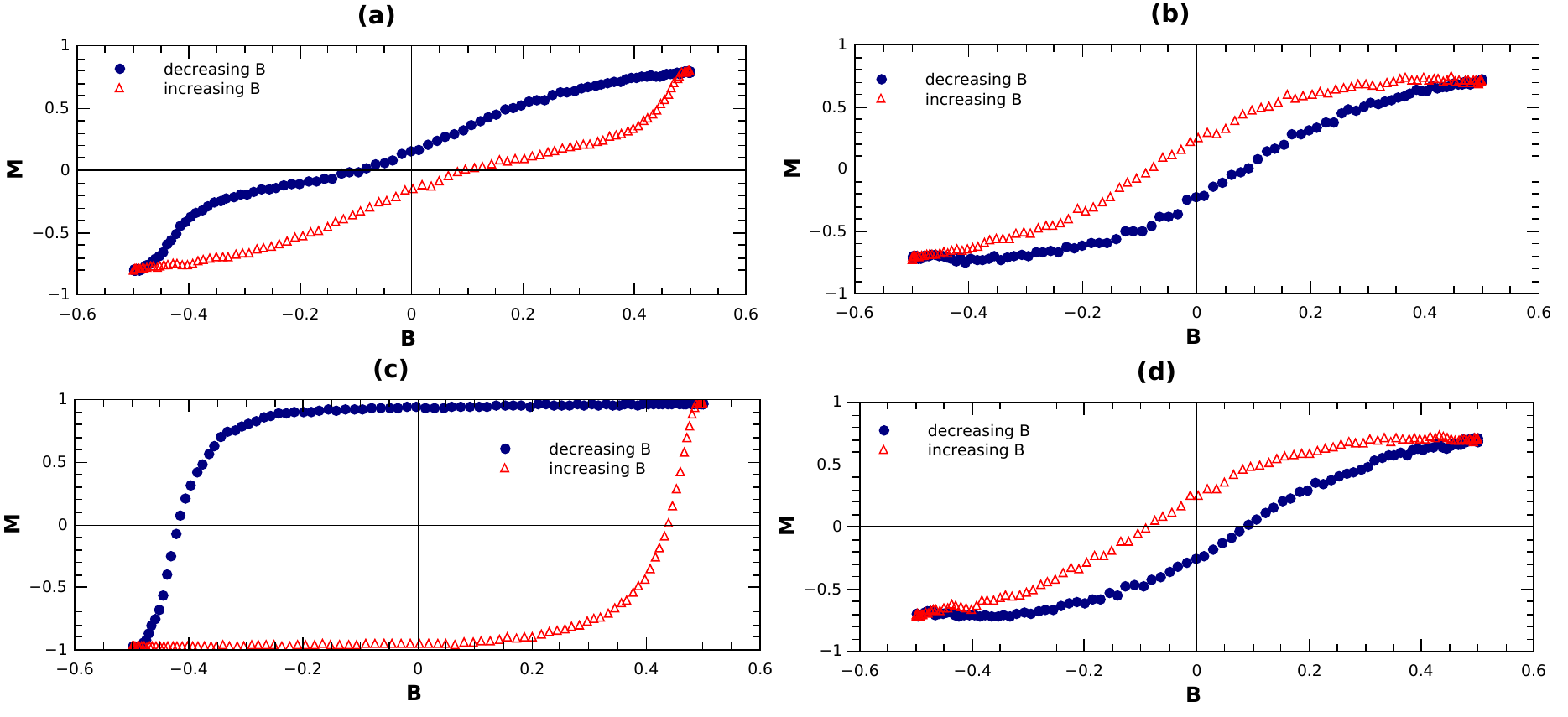}
		\caption[fig 10]{(a) Dynamic hysteresis loop of the trilayered system at $T=1.8$  for $B_0=0.5$ and $f=100$. Half of a cycle with decreasing $B$ and that for increasing $B$ are shown with separate symbols (see legend).  The same for the individual side layers are shown in (b) \& (d) and for middle one in (c).}
		\label{fig:fig10}
	\end{figure}
	
	On the other hand, within lower temperature regions (II) and (III) for small value of $ B_0$ (say at $ 0.5 $), $M$ of middle layer remains constant, while that of side layers changes to some extent and forms an asymmetric loop, making the loop for the trilayer highly asymmetric (see figure~\ref{fig:fig9}); while for same value of $ B_0 $ at temperature region (I), though the loops are symmetric,  due to different response to the oscillating external field, the hysteresis loop for whole system becomes highly warped (figure~\ref{fig:fig10}).
	In the last case, the contrast in responses is so prominent that while for the middle layer $M$ is lagging in phase behind the external field, for the side layers it is leading in phase (traversing the whole loop along clockwise direction instead of in anticlockwise way, which is usual for a ferromagnetic hysteresis loop). 
	
\begin{figure}[h!]
		\centering
		\includegraphics[width=.8\linewidth]{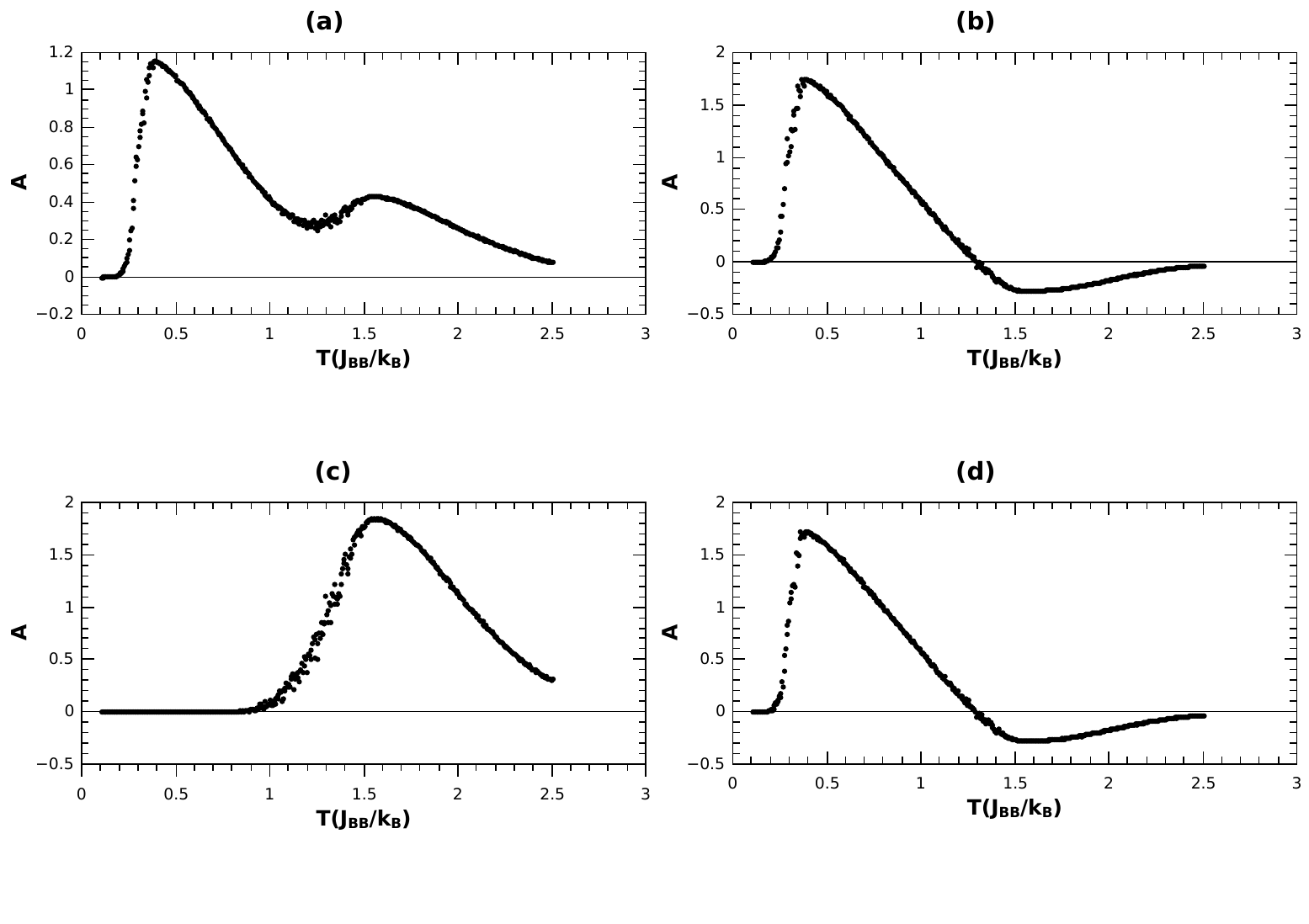}
		\caption[fig 11]{(a) Variation of $M-B$ loop area with temperature for B=0.5 and f=100. The contributions of individual side layers are shown in (b) \& (d), where that of the middle layer is depicted on (c). }
		\label{fig:fig11}
	\end{figure}

	At higher temperature the spin ordering gets weakened due to increase in thermal fluctuations. Hence $M$ varies in unison with the external time varying field and hysteresis loop area progressively shrinks. In 1993, Acharyya \& Chakrabarti studied the variation of loop area ($ A $) with temperature, frequency and amplitude of the external field for a bulk Ising ferromagnet~\cite{Acharyya1993} and they found monotonically decreasing nature of the loop area  with increase of temperature, within a chosen interval of $ T $ at a fixed $ B_0 $ and $ f $ of the applied field. But for this case though the general trend is deceasing it is not monotonic (figure~\ref{fig:fig11}). From the dynamic hysteresis loop of individual layers, it is evident that, at very low temperature the area of the loops vanish. Possibly because of low thermal excitations, spins of all the layers get frozen throughout the cycle. On the other hand at high temperature the layers become essentially paramagnetic, vanishing the dynamic hysteresis loop area. Hence there lies the maximum of the $ A-T $ curve in intermediate temperature region for each layer. But the position of the maximum corresponding to each layer, must be different due to difference in respective intra-layer interaction strengths. Moreover the peak region for the middle layer is associated with the temperature region where a minimum of loop area for the side layers lies. It is this temperature region where the clockwise orientation (mentioned in the previous paragraph) of the hysteresis loop of the side layers is observed; providing negative contribution to the hysteresis area of whole system (figure~\ref{fig:fig11}(b), (c), (d)). Thus the second peak in $ A-T $ curve is much smaller; resulting into non-monotonic but over all diminishing trend of the same at higher temperature zone (see figure~\ref{fig:fig11}(a)).
	
\begin{figure}[h!]
		\centering
		\includegraphics[width=0.8\linewidth]{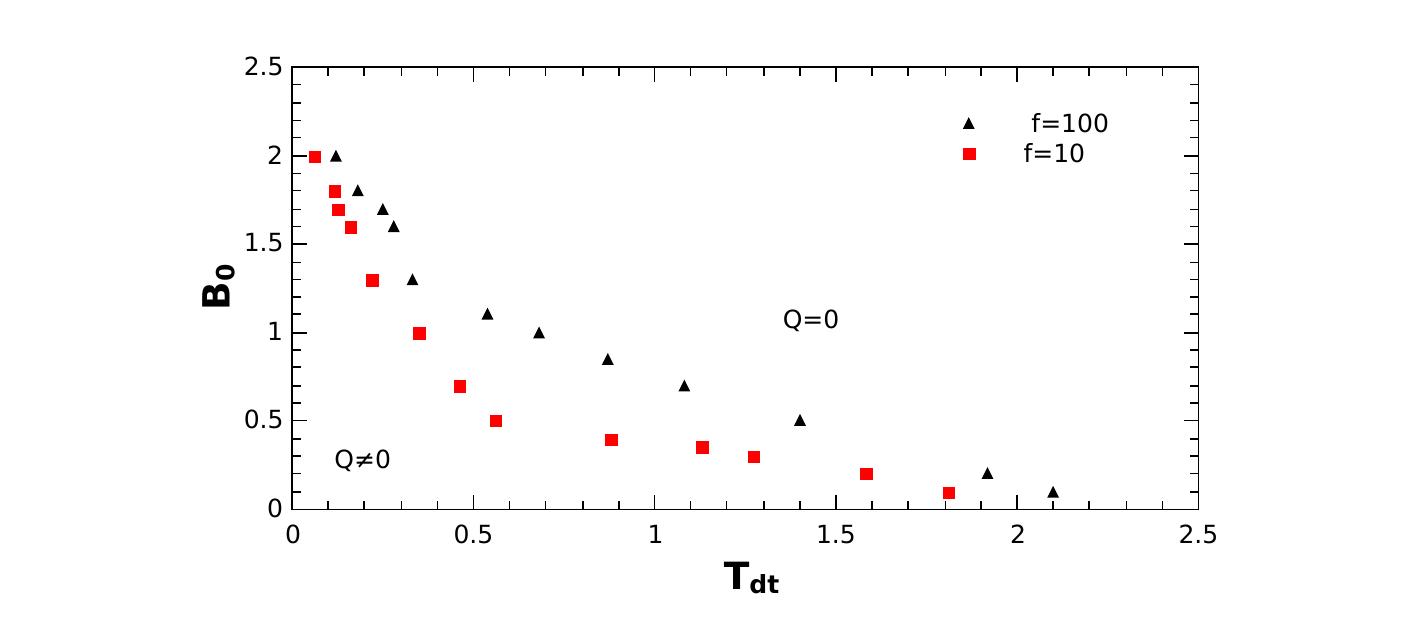}
		\caption[fig 12]{Phase boundaries in $B_{0}-T$ plane for $ f=100 $ and $ f=10 $ (see legends).}
		\label{fig:fig12}
	\end{figure}

	As the increasing thermal fluctuations eradicates the spin ordering, when temperature exceeds $T_{dt} $ for  fixed values of  $ B_0 $ and $ f $, the dynamically disordered phase emerges favouring the situation with $ Q=0 $. With the increase of $ B_0 $, as also reported  in the case of bulk Ising ferromagnets~\cite{Acharyya1993,Acharyya1995}, in this system too $T_{dt} $ gets shifted towards lower end; in $ B_0-T $ plane a phase boundary is found for a fixed $ f $ (figure~\ref{fig:fig12}). It is also found that with increase of $ f $, the phase boundary gets shifted towards higher end; which was also reported earlier for the case of bulk Ising ferromagnet~\cite{Acharyya1993,Acharyya1995}.
	
\begin{figure}[h]
		\centering
		\includegraphics[width=0.7\linewidth]{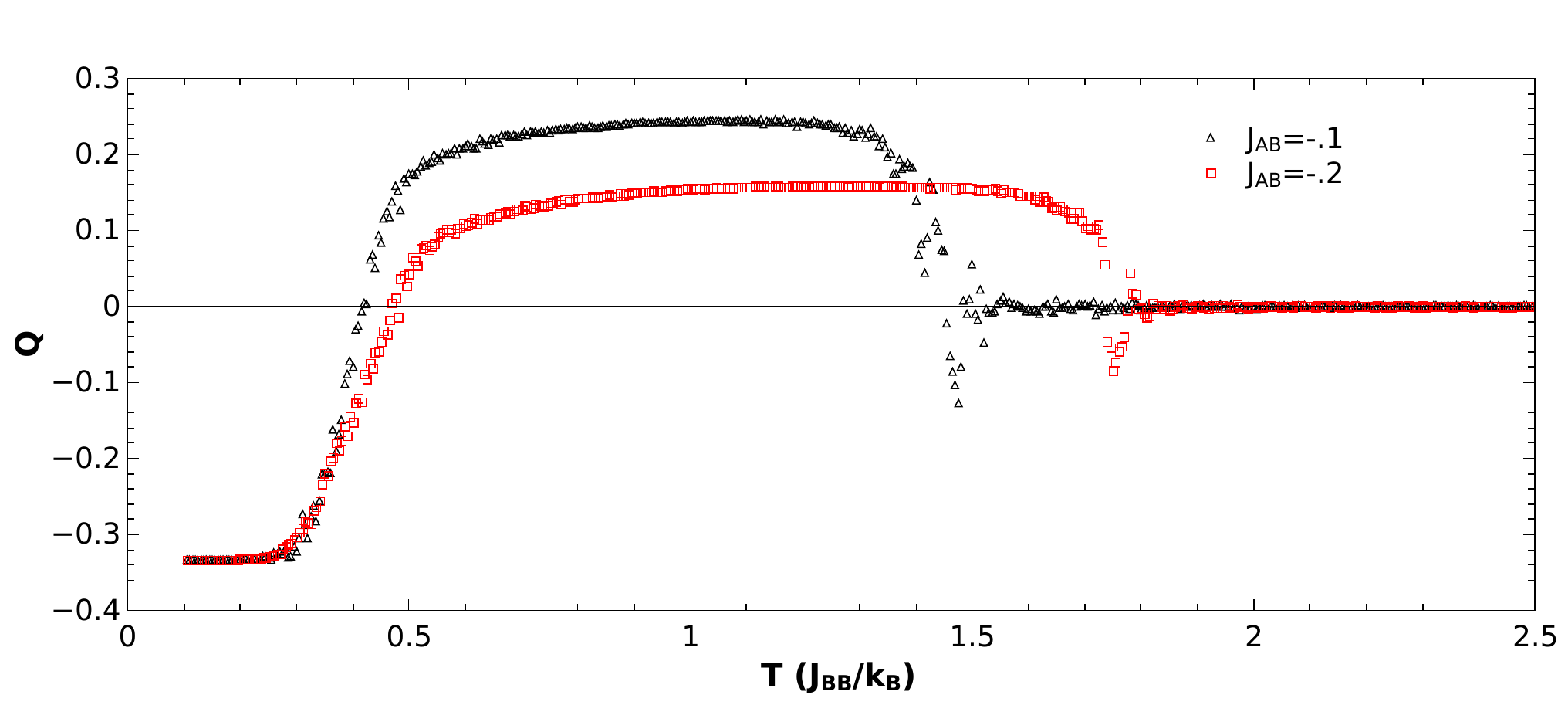}
		\caption[fig 13]{ Temperature variation of $ Q $ at two different values of $ J_{AB} $, and shifting of $ T_{dt} $.}
		\label{fig:fig13}
	\end{figure}

	But in this trilayered system there is a role of AFM inter-layer coupling too. Already in this article, it has been discussed that ordering in the middle layer (type {\bf B}) is much more robust compared to side layers (type {\bf A}). $ Q $ ultimately vanishes at a temperature where $ Q $ for the middle layer becomes zero; but it is also observed that in this trilayered system the AFM interlayer coupling, however weak in magnitude, also plays a significant role. With the increase of $ J_{AB} $, the spin ordering in middle layer and side layers reinforce each other to push the transition temperature $ T_{dt} $ towards higher end (figure~\ref{fig:fig13}). It is an interesting feature in DPT, characteristic to
	these types of trilayered materials only.  
	
	\section{Concluding Remarks}
	In this work DPT with time varying harmonic external field has been observed in ${\bf ABA}$ trilayered structure. Apart from DPT, which is found for bulk system too, dynamic compensation emerges as a unique feature for such type of trilayered system. But it should be kept in mind that  here a classical Ising Hamiltonian is used. At very low temperature (i.e when $k_BT$ is negligible compared to the spin-spin average interaction energy) this may not reliably portray the realistic situation.
	
	The unique signature of this type of system in dynamic response is also observed in warping of the dynamic hysteresis loop; which has also been discussed at length. The non-monotonic variation of loop area with temperature is found to be another reflection of the peculiarity of these trilayered systems. It has also been shown that even a weak AFM inter-layer interaction plays a significant role in determining transition temperature; this may be considered to be an important handle in DPT in this kind of systems. Here the time dependent external magnetic field is chosen to be harmonic. It may be interesting to investigate the effect of other forms of time dependence for the magnetic field.  
	
	%
\end{document}